%Paper: gr-qc/9505037
%From: lsk@iucaa.ernet.in (L. Sriramkumar)
%Date: Sat, 20 May 95 16:08:59+050

%The document begins here%%%%%%%%%%%%%%%%%%%%%%%%%%%%%%%%%%%%%%%%%%%%%%%%%
\magnification=\magstep1
\tolerance=2000

\font\large=cmbx10 at 12 pt
\newcount\equationno      \equationno=0
\newtoks\chapterno \xdef\chapterno{}
\def\eqn{\eqno\eqname}
\def\eqname#1{\global \advance \equationno by 1 \relax
\xdef#1{{\noexpand{\rm}(\chapterno\number\equationno)}}#1}

\centerline{\large Limits on the validity of the}
\medskip
\centerline{\large semiclassical theory}
\vskip 1truein
\centerline{\bf L. Sriramkumar$^{\dagger}$}
\centerline{IUCAA, Post Bag 4, Ganeshkhind,}
\centerline{Pune 411007, INDIA.}
\vskip 1truein
\centerline{\bf Abstract}
\bigskip
\noindent
For want of a more natural proposal, it is generally assumed that the
back-reaction of a quantised matter field on a classical metric is
given by the expectation value of its energy-momentum tensor, evaluated
in a specified state.
This proposal can be expected to be quite sound only when the fluctuations
in the energy-momentum tensor of the quantum field are negligible.
Based on this condition, a dimensionless criterion has been suggested earlier
by Kuo and Ford for drawing the limits on the validity of this semiclassical
theory.
In this paper, we examine this criterion for the case of a toy model,
constructed with two degrees of freedom and a coupling between them that
exactly mimics the behaviour of a scalar field in a Friedmann universe.
To reproduce the semiclassical regime of the field theory, in the toy model,
one of degrees of freedom is assumed to be classical and the other quantum
mechanical.
Also the backreaction is assumed to be given by the expectation values of
the quantum operators involved in the equations of motion for the classical
system.
Motivated by the same physical reasoning as Kuo and Ford, we, here, suggest
another criterion, one which will be shown to perform more reliably as we
evaluate these criterions for different states of the quantum system in the
toy model.
Finally, from the results obtained we conclude that the semiclassical theory
being considered for the toy model is reliable, during all stages of its
evolution, only if the quantum system is specified to be in coherent like
states.
The implications of these investigations on field theory are discussed.
\vskip 0.6 true in
\centerline{{\bf IUCAA -- 15/95, April, 1995} : Submitted for publication}
\vfill
\hrule
\bigskip
$^{\dagger}$ sriram@iucaa.ernet.in
\vfill\eject

\beginsection{\bf 1. Introduction}

There exists a domain during the evolution of the universe when
the energies of the ongoing physical processes lie between Compton
and Planck scales.
In this domain, though it is sufficient to describe gravity by a
classical metric the quantum nature of any matter field present
has to be taken into account.
In general relativity, the theory which we assume to describe
gravity adequately in the regime of our interest, it is the
energy-momentum tensor of the matter field, $T_{\mu\nu}$,
that is reponsible for the classical geometry.
The energy-momentum tensor for a quantum field being an operator,
a c-number ought to be constructed out of this operator
before the effect of the quantum field on a classical metric can
be studied.
It has been suggested earlier in literature$^{[1]}$, that the transition
element $\langle\, out\, \vert\, {\hat T}_{\mu\nu}\,\vert\, in\rangle$
(where $\vert\, in\, \rangle$ and $\vert\, out \, \rangle$ are the
asymptotic states of the quantum field), obtained by the variation
of the effective action, be considered as the backreaction term.
This transition element is in general a complex quantity and may lead
to a complex metric which will prove rather difficult to interpret
unless the imaginary part happens to be negligible or is dropped
in an ad hoc manner.
A more natural and plausible proposal$^{[2,\, 3,\, 4]}$ would be to
consider the expectation value of the energy-momentum operator of the
quantum field as the term that induces the non-trivial geometry.
Since the theory being considered here, by itself, is
incapable of providing us with a preferred state for the quantum
matter field, the expectation values are to be evaluated in a
state that has to be specified by hand.
So the analysis of the action of a quantum field; say a
massless scalar field, on the classical background metric
reduces to that of solving the Einstein's equations

$$G_{\mu\nu}\; =\; R_{\mu\nu}\; -\; {1 \over 2}\, g_{\mu\nu}\, R\;
=\; 8\, \pi\, \big\langle\,{\hat T}_{\mu\nu}\,\big\rangle\eqn\qee$$
where $\big\langle\,{\hat T}_{\mu\nu}\,\big\rangle$ is the expectation
value of the energy-momentum operator (in the specified state)
and the Klein-Gordon equation

$$\nabla_{\mu}\, \nabla^{\mu}\, {\hat {\Phi}}(x)\; =\; 0\eqn\qkg$$
where ${\hat {\Phi}}$ is the operator corresponding to the quantised
scalar field, self-consistently. (We adopt the convention
${\hbar}=\, G=\, c=\,1$ and a metric signature of $(-2)$ in this paper.)
\smallskip
Apart from the fact that the geometry has to be classical for the
semiclassical theory proposed above to be reliable, {\it i.e} the
energy scales involved should be far below the Planck scale, the
fluctuations in the energy-momentum densities of the quantum field
should not be too large either$^{[3]}$,
{\it i.e}

$$\big\langle\,{\hat T}_{\alpha \beta}(x)\, {\hat T}_{\mu\nu}(y)\,\big
\rangle\; \approx \; \big\langle\,{\hat T}_{\alpha \beta}(x)\,\big\rangle
\, \big\langle\,{\hat T}_{\mu\nu}(y)\,\big\rangle.\eqn\qfl$$
So, the semiclassical theory based on {\qee} cannot be trusted for
those states of the quantum field where the fluctuations in the
energy-momentum densities are too large.
The goal of this present paper is to check the validity of the
semiclassical theory based on the equations {\qee} and {\qkg}
in time dependant background metrics like for instance, Friedmann
models, for different states prescribed for the quantum field.
\smallskip
The calculations necessary for drawing the limits, with aid of the
constraint equation {\qfl}, on the validity of the semiclassical
theory proposed above will involve divergences of quantum field theory
and their regularisation procedures.
Since these schemes might eventually prove to sidetrack the issue of
our concern, instead of analysing the validity of the semiclassical theory
for the case of quantised scalar fields in time dependant metrics, we,
in this paper, will study the same for the case of a toy model.
\smallskip
The toy model will consist of two degrees of freedom.
It will be constructed such that the coupling between the two degrees
of freedom exactly resembles that of a scalar field evolving in a
Friedmann universe.
Of the two degrees of freedom, one of them will be assumed to be
classical and the other quantum mechanical so that the toy model
reproduces the semiclassical nature of the field theory in the
domain of our interest.
The analysis of the the validity of the semiclassical theory will be
carried out for the toy model with the assumption that the backreaction
term is given by the expectation values (in the specified state) of
the quantum operators involved in the equations of motion for the
classical degree of freedom.
\smallskip
This paper is organised as follows.
In section {\bf 2} we construct the action for the toy model taking cues
from the coupling of the scalar field to a Friedmann metric, obtain the
classical equations of motion and then extend the relevant equations to
the semiclassical domain as described in the previous paragraph.
In section {\bf 3}, the quantisation of a time dependant simple harmonic
oscillator, the quantum degree of freedom in the toy model, is carried
out in the Heisenberg picture.
In section {\bf 4} we briefly review the motivations behind the criterion
proposed by Kuo and Ford to draw the limits on the validity of the
semiclassical theory, evaluate this criterion for a particular case
in the toy model and point out its drawbacks.
We then suggest another criterion, based on the same physical reasoning
as Kuo and Ford's, but one that is more reliable, to obtain the limits
on the validity of the semiclassical theory being considered for the
toy model.
In section {\bf 5} we calculate the criterion we have suggested and
the one that has been put forward by Kuo and Ford for three different
quantum states of the simple harmonic oscillator {\it viz} (i) vacuum,
(ii) $n$th excited and (iii) coherent states.
In the final section {\bf 6}, we comment on the conclusions that can
be drawn from our investigations and also disuss the possible implications
of our analysis on the study of quantum field theory in time dependant
background metrics.
\bigskip

\beginsection{\bf 2. Constructing the action for the toy model}

The action for a massless scalar field coupled to gravity is given by
the equation

$${\cal A}\; =\; \int d^4x\, \sqrt{-g}\, \left(\,{1 \over 16\,\pi}\,
R\; +\; {1 \over 2}\,\partial_{\mu}\Phi\,
\partial^{\mu}\Phi \, \right).\eqn\qact$$
Variation of the action {\qact} with respect to the metric $g_{\mu\nu}$
and the field $\Phi$ would lead us to the Euler-Lagrange equations for
gravity and the matter field repsectively ({\it viz} equations {\qee}
and {\qkg} with just the classical $\Phi$ and $T_{\mu\nu}$ and not
their quantum operators).
For a spatially flat Friedmann model described by the metric

$$ds^2\; =\; dt^2\; -\; a^2(t)\, \left(\,dx^2\; +\; dy^2\; +\; dz^2\,
\right)
\eqn\qmet$$
where $a$ is the scale factor, the action {\qact} reduces to be

$${\cal A}\; =\; \int d^3x\, dt\, a^3 \left\lbrace\,-{3 \over {8\,
\pi}}\, \left(\,{\ddot a \over a}\; +\; {{\dot a}^2 \over {a^2}}
\,\right)\; +\; {1 \over 2}\, \left(\,\left\lbrace\,{\partial \Phi
\over {\partial t}}\right\rbrace^2\; - \; \left\lbrace\,{\vert
\nabla \Phi \vert \over a}\right\rbrace^2\,\right)\,\right\rbrace\;
\eqn\qq$$
where the dots denote derivatives with respect to the comoving
time $t$.
Since any action is expected to involve derivatives of the degrees of
freedom no further than the first, the terms involving the second time
derivatives of $a$ in the above action can be eliminated by integrating
them by parts with respect to $t$.
The reduced action, after the integration by parts, is obtained to be

$${\cal A}\; =\; \int d^3x\, dt\, a^3 \left\lbrace\, -{3 \over {8 \pi}}\,
\left(\, {{\dot a}^2 \over {a^2}}\,\right)\; +\; {1 \over 2}\,
\left(\,\left\lbrace\,{\partial \Phi \over {\partial t}}\,\right\rbrace^2\;
- \; {\left\lbrace\,{\vert \nabla \Phi \vert \over a}\,\right\rbrace}^2\,
\right)\,\right\rbrace.\eqn\qract$$
\smallskip
For a time dependant metric, assumed to be homogenous and isotropic,
varying the action {\qact} with respect to the components $g_{00}$ and
$g_{ik}(i,\,k\, =\,1,\,3)$ of the metric tensor ($g_{0i}\,=\,0$ due to
isotropy) yields the two Friedmann equations.
Of these two, the equation obtained by varying $g_{ik}$ involves
the second time derivatives of the scale factor, whereas the
one obtained by varying $g_{00}$ is a constraint equation and
depends only the on first time derivatives of $a$.
On the other hand, the variation of the action {\qract} with respect
to $a$ will only yield the equation involving the second time
derivatives of $a$ and constraint equation can not be obtained from the
reduced action.
The reason being that in arriving at {\qract} from {\qact} we have lost
the time-reparametrisation invariance of a Friedmann model having chosen
$g_{00}$ to be unity in the metric {\qmet}.
To reproduce this `lost' equation, the relevant degree of freedom has
to be re-introduced in the Friedmann metric.
Introducing an arbitrary function $N(t)$ into the metric {\qmet}, as
follows

$$ds^2\; =\; N^2(t)\,dt^2\; -\; a^2(t)\, \left(\,dx^2\; +\; dy^2\;
+\; dz^2\,\right),\eqn\qNmet$$
we obtain the reduced action for the above metric to be

$${\cal A}\; =\; \int d^3x\, dt\, a^3 \left\lbrace\,-{3 \over
{8\, \pi\, N(t)}}\,\left(\,{{\dot a}^2 \over {a^2}}\,\right)\; +\; {1 \over 2}
\,\left(\,{1 \over {{N(t)}}}\,\left\lbrace\,{\partial \Phi \over {\partial
t}}\,\right\rbrace^2\;- \; N(t)\,\left\lbrace\,{\vert \nabla \Phi
\vert \over a}\right\rbrace^2\,\right)\,\right\rbrace.\eqn\qractN$$
Now, varying the above action with respect to $N$ and setting $N$
to be unity after the variation yields

$$\left(\,{{\dot a}^2 \over {a^2}}\,\right)\; =\; {4 \; \pi \over 3}\,
\left\lbrace\,\left(\,{\partial \Phi \over {\partial t}}\,\right)^2\;
+ \; \left(\,{\vert \nabla \Phi \vert\over a}\,\right)^2\,\right\rbrace
\eqn\qfr$$
which is the Friedmann equation we are interested in.
\smallskip
Having learned the aspects of the coupling of a scalar field to a
Friedmann metric from the brief review above, it is easy to see that the
following action for two coupled degrees of freedom $C$ and $q$

$${\cal A}\; =\; \int dt\, \left\lbrace\,-{M \over 2}\,
{{\dot C}^2 \over N}\; +\; {m \over 2}\, {{\dot q}^2 \over  N}\;
-\; N\, {m \over 2}\, \omega^2(C)\, q^2 \,\right\rbrace\eqn\qmact$$
exactly reproduces all the features of our interest in the reduced action
{\qract}.
(The dots hereafter denote derivatives with respect to $t$, the time
parameter for the toy model.)
The variation of the action {\qmact} with respect
to $N$ when $N$ is set to be unity after the variation, yields the
equation of constraint

$$ {M \, {\dot C}^2 \over 2}\;=\; \left(\,{m\, {\dot q}^2 \over 2}\,
+\; {m\, \omega^2(C)\, q^2 \over 2}\,\right).\eqn\qq$$
The above equation can be expressed as

$${M {\dot C}^2 \over 2}\;=\; H\eqn\qCc$$
where $H$, the Hamiltonian corresponding to the variable $q$ is
given by

$$H\; =\; \left(\,{p^2 \over {2m}}\; +\; {m\, \omega^2(C)\,
 q^2 \over 2}\,\right).\eqn\qH$$
($p$ in the above equation represents the momentum conjugate to the
degree of freedom $q$), similar in structure to the Friedmann equation
{\qfr}. The Euler-Lagrange equation for the variable $q$ is

$$\ddot q\; +\; \omega^2(C)\, q\; =\; 0.\eqn\qqc$$
By comparing the actions {\qact} and {\qmact}, it can be easily seen that
the degree of freedom $C$ is expected to behave like the scale factor $a$
of a Friedmann model and the the variable $q$ is to mimic the behaviour
of the scalar field.
\smallskip
Also, if our toy model is to reproduce the behaviour of a {\it quantised}
scalar field in a Friedmann metric, we have to assume that the degree of
freedom $q$ behaves quantum mechanically.
Then extending the classical equations of motion for $C$ and $q$ to
the semiclassical domain in a fashion similar to what has been done
in {\qee}, we obtain that

$${M \, {\dot C}^2 \over 2}\;=\; \big\langle\,{\hat H}(t)\,\big\rangle\;
=\; E(t)
\eqn\qCsc$$
where $\big\langle\,{\hat H}(t)\,\big\rangle$ is the expectation value (in
the specified state) of the Hamiltonian operator corresponding to the
classical Hamiltonian given by {\qH} for the degree of freedom $q$.
\smallskip
The evolution of quantum system $q$, a time-dependant oscillator will
be discussed in the following section.
\bigskip

\beginsection{3. Quantisation of the time dependant oscillator:
Heisenberg picture}

The Hamiltonian operator corresponding to the the degree of freedom $q$
is

$$\hat H\; =\; {{\hat p}^2 \over 2m}\; +\; {m \over 2}\,
\omega^2(C)\, {\hat q}^2\eqn\qHop$$
where ${\hat q}$ and ${\hat p}$ are the operators corresponding to the
variable $q$ and its conjugate momentum $p$.
In the Heisenberg picture the operators are dependant on time and
the operators ${\hat q}$ and ${\hat p}$ satisfy the equations$^{[5]}$

$${d{\hat q} \over {dt}}\; =\; i\, [\hat H,\, \hat q]\; =\; {{\hat p}
\over m}\eqn\qqop$$
and

$${d{\hat p} \over {dt}}\; =\; i\, [\hat H,\, \hat p]\;=\; -\, m\,
\omega^2(C)\, {\hat q}.\eqn\qpop$$
Substituting {\qqop} in {\qpop} we obtain that

$${d^2 {\hat q} \over dt^2}\; +\; \omega^2(C)\, {\hat q}\; =\; 0\eqn\qqqop$$
the operator equation corresponding to the classical equation {\qqc}.
\smallskip
The conjugate variables $q$ and $p$ being observables, {\it viz} the
position and the momentum of the harmonic oscillator, the
corresponding operators are to be hermitian, {\it i.e} ${\hat q}
^{\dagger}\,=\,{\hat q}$ and ${\hat p}^{\dagger}\,=\,{\hat p}$.
So the solutions to the Heisenberg equations of motion {\qqop}
and {\qpop} can be written down to be

$${\hat q}\; =\; \left(\,{\hat A}\, Q\; +\; {\hat A}^{\dagger}\,
Q^*\,\right)\eqn\qqopAQ$$
and

$${\hat p}\; =\; m \, {d{\hat q} \over {dt}}\;= \; m\, \left(\,{\hat A}\,
{\dot Q}\; +\; {\hat A}^{\dagger}\, {\dot Q}^*\,\right),\eqn\qpopAQ$$
where ${\hat A}$ is a time-independant operator and $Q$ satisfies the
differential equation

$${d^2 Q \over dt^2}\; +\; \omega^2(C)\, Q\; =\; 0.\eqn\qQ$$
This equation can be solved by the ansatz$^{[6]}$

$$Q\; =\; \left(\,{\alpha(t)\, f\; +\; \beta(t)\, f^* \over
{\sqrt{2\,m\,\omega}}}\,\right)\eqn\qQansz$$
where

$$f\; =\; exp-\,\left\lbrace\,\int_{t_0}^t dt'\, \omega\left\lbrace\, C(t')
\,\right\rbrace\,\right\rbrace\eqn\qf$$
and $t_0$ is an {\it early} time when the initial conditions for the
differential equation {\qQ} will be specified.
Defining ${\dot Q}\,=\,(dQ/dt)$ to be

$$\dot Q\; =\; (-i\; \omega)\, \left(\,{\alpha\, f\; +\; \beta\; f^*
\over {\sqrt{2\,m\,\omega}}}\,\right)\eqn\qQdot$$
(where $f^*$ is the complex conjugate of $f$) and incorporating {\qQansz}
and {\qQdot} in {\qQ} we find that $\alpha$ and $\beta$ satisfy the set
of coupled equations

$$\dot {\alpha}\; =\; \left(\,{\dot {\omega} \over {2\,\omega}}\,\right)\,
\beta\, {f^*}^2\quad ;\quad\dot {\beta}\; =\; \left(\,{\dot {\omega}
\over {2\,\omega}}\,\right)\, \alpha\, f^2.\eqn\qalpbet$$
Integrating the wronskian condition for the differential equation {\qQ},
{\it viz}

$${d \over {dt}}\left(\,Q\, {\dot Q}^*\; -\; Q^*\, \dot Q\,\right)\; =\; 0$$
we obtain that

$$\left(\,Q\, {\dot Q}^*\; -\; Q^*\, \dot Q\,\right)\; =\; {i \over m}
\eqn\qwronsk$$
where the constant of integration has been chosen to be $(i/m)$.
Substituting {\qQansz} and {\qQdot} into the above equation we find
that the wronskian condition reduces to the relation

$${\vert\, \alpha\, \vert}^2\; -\; {\vert\, \beta\, \vert}^2\; =\;
1.\eqn\qalpbetwr$$
Also, when {\qQansz} and {\qQdot} are substituted  in the equations
{\qqop} and {\qpop} we obtain that the evolution of the two operators
$\hat q$ and $\hat p$ are described by the equations

$$\hat q\; =\; {1 \over {\sqrt{2\, m\,\omega}}}\,
\biggl\lbrace\, \left(\,\alpha\,
{\hat A}\; +\; {\beta}^*\, {\hat A}^{\dagger}\,\right)\; f\;+\; \left(\,
{\alpha}^*\, {\hat A}^{\dagger}\; +\; \beta\; {\hat A}\,\right)f^*\,\biggl
\rbrace\eqn\qqopsoln$$
and

$$\hat p\; =\; i\, \sqrt{m\, \omega \over 2}\; \biggl\lbrace \,-\left
(\,\alpha\, {\hat A}\; +\; {\beta}^*\, {\hat A}^{\dagger}\,\right)\, f\;
+\; \left(\,{\alpha}^*\, {\hat A}^{\dagger}\; +\; \beta\, {\hat A}\,
\right)f^*\,\biggl\rbrace.\eqn\qpopsoln$$
When the above solutions are substituted in the following commutation
relations for the operators corresponding to the canonically conjugate
variables

$$[{\hat q},\, {\hat q}]\; =\; 0\quad; \quad[{\hat p},\,{\hat p}]\quad
;\quad[{\hat q},\, {\hat p}]\;=\;i, \eqn\qqpc$$
we obtain the commutation relations between the operators
${\hat A}$ and ${\hat A}^{\dagger}$ to be

$$[{\hat A},\, {\hat A}]\; =\; 0 \quad; \quad[{\hat A}^{\dagger},\,
{\hat A}^{\dagger}]\; = \;0\quad ; \quad [{\hat A},\, {\hat A}^{\dagger}]\;
=\;1.\eqn\qAc$$
If the initial conditions for the equations of motion are chosen to be
such that $\alpha(t_0)\,=\,1,\; \beta(t_0)\,=\,0$ and
$\omega(t_0)\,=\,\omega_0$ (an arbitrary constant) at $t\, =\, t_0$ then
the operators ${\hat q}$ and ${\hat p}$ at $t\,=\,t_0$ are given by the
equations

$${\hat q}\; =\; {1 \over \sqrt{2\,m\,\omega_0}}\, \left(\,{\hat A}\; +\;
{\hat A}^{\dagger}\,\right)\quad; \quad{\hat p}\; =\; i\,\sqrt{m\, \omega_0
\over 2}\, \left(\,-{\hat A}\; +\; {\hat A}^{\dagger}\,\right).\eqn\qqptnot$$
\smallskip
In the Heisenberg picture the quantum states are independent of time.
The state of the quantum system is usually prescribed at the same instant
when the initial conditions for the equations of motion are specified.
The quantum states for the matter field in which the backreaction problem
in field theory is generally studied are the vacuum, $n$-particle and
coherent states.
The corresponding states for the quantum oscillator in our toy model
can be defined as follows: the vacuum state $\vert\, 0\,\rangle$ satisfies
the conditon

$${\hat A}\, \vert\, 0\,\rangle\;=\;0,$$
a $n$-particle state $\vert\, n\,\rangle$ is defined such that

$${\hat A}^{\dagger}\,{\hat A}\, \vert\, n\,\rangle\; =\; n\,
\vert\, n\, \rangle$$
and a coherent state $\vert\, \lambda\,\rangle$ follows the equation

$${\hat A}\, \vert\, \lambda\, \rangle\; =\; \lambda\,
\vert\, \lambda\,\rangle.$$
Substituting {\qqopsoln} and {\qpopsoln} in {\qHop} we can see that the
Hamiltonian operator at any time $t\,>\,t_0$ is given by

$${\hat H}(t)\; =\; \left\lbrace\,{\hat a}^{\dagger}(t)\,{\hat a}(t)\;
+\; {1 \over 2}\,\right\rbrace\, \omega \eqn\qHopt$$
where

$${\hat a}(t)\; =\; \alpha(t)\, {\hat A}\; +\; {\beta}^*(t)\, {\hat A}^
{\dagger}.\eqn\qa$$
The equation of motion for $C$, {\qCsc}, then reduces to

$${M\, {\dot C}^2 \over 2}\; =\; \big\langle\, {\hat H}(t)\,\big\rangle
\;= E(t)\; =\; \left\langle\,\left\lbrace\,{\hat
{{\hat a}^{\dagger}}}(t)\,{\hat a}(t)\; +\;
 {1 \over 2}\,\right\rbrace\,\omega\right\rangle
\eqn\qCscalpbet$$
where the expectation value is evaluated in the state specified
for the quantum oscillator.
The expectation value of the square of the Hamiltonian operator,
which will be needed later to evaluate the fluctuations in the energy,
is

$$E^2(t)\;=\;\big\langle\,{\hat H}^2(t)\,\big\rangle\;=\;\biggl\langle\,
\left\lbrace\,{\hat{{\hat a}^{\dagger}}}(t)\,{\hat a}(t)\; +\;
{1 \over 2}\,\right\rbrace\,\omega\,\left\lbrace\,
{\hat a}^{\dagger}(t)\,{\hat a}(t)\; +\;{1 \over 2}\,\right\rbrace
\, \omega\biggl\rangle.\eqn\qEtwo$$
\bigskip

\beginsection{4. Criterions for drawing the limits on
semiclassical theory}

The semiclassical theory as described by the equations {\qee} and
{\qkg} does not account for the fluctuations in the energy-momentum
densities of the quantum field.
So, as mentioned earlier, this theory can be relied upon only when
the fluctuations in the energy-momentum densities are small when
compared to their expectation values.
\smallskip
Motivated by this fact, Kuo and Ford$^{[7]}$ have suggested that the
dimensionless quantity

$$\Delta_{\alpha\beta\mu\nu}(x, \;y)\;\equiv\; \biggl\vert\,
{\big\langle\,:{\hat T}_{\alpha\beta}(x)\,{\hat T}_{\mu\nu}(y):\,
\big\rangle\;
-\; \big\langle\,:{\hat T}_{\alpha\beta}(x):\,\big\rangle\,
\big\langle\,:{\hat T}_{\mu\nu}(x):\,\big\rangle \over
{\big\langle\,:{\hat T}_{\alpha\beta(x)}\,
{\hat T}_{\mu\nu}(y):\,\big\rangle}}\,\biggl\vert\eqn\qdxy$$
(where the semicolons represent renormalisation of the expectation
values) be considered as a measure of the fluctuations in the
energy-momentum densities of the quantum field.
When the fluctuations in the energy-momentum densities are negligible,
this quantity will be far less than unity and the semiclassical theory
being considered here will prove to be quite sound.
But when the fluctuations are large the above quantity is expected
to be of order unity reflecting a complete breakdown of the proposed
semiclassical theory.
\smallskip
The numerous components and the dependance on the two spacetime points
make the quantity $\Delta_{\alpha\beta\mu\nu}(x, \,y)$ an extremely
cumbersome object to handle.
For the sake of simplicty, as Kuo and Ford themselves suggest,
we can confine our attention to either the evaluation of the purely
temporal component of this quantity in the coincidence limit ({\it i.e}
when $x\, \rightarrow\, y$)

$$\Delta_{KF2}(x)\;\equiv\;
\biggl\vert\,{\big\langle\,:{\hat T}_{00}^2(x):\,\big\rangle\;
-\; {\big\langle\,:{\hat T}_{00}(x):\,\big\rangle}^2 \over
{{\big\langle\,:{\hat T}_{00}^2(x)}
:\,\big\rangle}}\,\biggl\vert\eqn\qFtwo$$
(subscript $KF$ standing for Kuo and Ford) or the criterion

$$\Delta_{KF1}(x)\;\equiv\;\biggl\vert{\big\langle\,:
{\hat T}_{00}^2(x):\,\big\rangle\; -\; {\big\langle\,:
{\hat T}_{00}(x):\,\big\rangle}^2 \over {{\big\langle\,:
{\hat T}_{00}(x):\,\big\rangle}^2}}\,\biggl\vert.\eqn\qFone$$
The quantities $\Delta_{KF1}$ and $\Delta_{KF2}$ are related to each
other by the equation

$$\Delta_{KF2}\; =\; \left(\,{\Delta_{KF1} \over {\Delta_{KF1}\; +\; 1}}\,
\right).\eqn\qq$$
In the next section, when similar quantities are evaluated for
the case of our toy model it will be shown that the criterions
$\Delta_{KF1}$ and $\Delta_{KF2}$ yield equivalent results.
The evaluation of $\Delta_{KF2}$ was carried out by Kuo and Ford for
different states of a quantised massless scalar field in flat space.
The limits on the validity of the semiclassical theory can not be drawn
from the evaluation of these criterions in flat space but has to be
carried out for the case of quantum fields in curved backgrounds.
\smallskip
For the Friedmann model, the time dependant metric for which we are trying
to obtain the limits on the validity of the semiclassical theory as
described by equations {\qee} and {\qkg}, the adiabatic limit corresponds
to the case when the scale factor $a$ is a slowly varying function of time,
{\it i.e} when $\left(\,{\dot a}\,/\,a\,\right)\,\rightarrow\,0$,
In this limit the Friedmann metric is almost Minkoskian and in flat
space the fluctuations in the energy-momentum densities {\it are} neglible.
Extending this example, in the adiabatic limit, the fluctuations in
the energy-momentum densities of a quantum field in an arbitrary spacetime
can be, in general, expected to be small.
And a dimensionless criterion, supposed to reflect the magnitude of
these fluctuations, proposed for drawing the limits on the validity
of the semiclassical theory, should identically vanish in this limit.
\smallskip
The adiabatic limit for our toy model corresponds to the case when the
coupling term $\omega\left(C(t)\right)$ is a slowly evolving function
in time, {\it i.e} when $(d\omega/dt)\, \rightarrow\,0$.
The equation {\qalpbet} implies that, for the initial conditions that
have been specified ({\it viz} $\alpha(t_0)\,=\,1$ and $\beta(t_0)\,=\,0$),
in the adiabatic limit $\alpha\, \rightarrow\, 1$ and
$\beta\, \rightarrow\, 0$ for $t\,>\,t_0$.
And in this limit, since the semiclassical theory that has been put forward
for the toy model is known to be reliable, any criterion suggested for
drawing the limits on its validity should reduce to zero when
$\beta\,\rightarrow\,0$.
\smallskip
The back-reaction term (refer to {\qCsc}) for our toy mechanical model
is the expectation value of the Hamiltonian corresponding to the quantum
degree of freedom.
So, for the toy model, it is the magnitude of the fluctuations in the
energy of the quantum oscillator that will decide the validity of the
semiclassical theory.
So the quantities equivalent to ${\Delta}_{KF1}$ and ${\Delta}_{KF2}$ for
the case of our toy model are

$$\Delta_{KF1}(t)\;\equiv\;\biggl\vert\,{\big\langle\,:{\hat H}^2(t):\,\big
\rangle\; -\; {\big\langle\,:{\hat H}(t):\,\big\rangle}^2 \over
{{\big\langle\,:{\hat H}(t):\,\big\rangle}^2}}\,\biggl\vert,\eqn\qFonetm$$
and

$$\Delta_{KF2}(t)\;\equiv\; \biggl\vert\,{\big\langle\,:{\hat H}^2(t):\,\big
\rangle\; -\; {\big\langle\,:{\hat H}:\,\big\rangle}^2 \over
{{\big\langle\,:{\hat H}^2}(t):\,\big\rangle}}\,\biggl\vert.\eqn\qFtwotm$$
The semi-colons in the above quantities represent regularisation,
performed either by normal ordering of the operators or by vacuum
subtraction.
Renormalisation of the expectation values can be achieved through normal
ordering by moving all the ${\hat a}^{\dagger}$'s  to the left of
${\hat a}$'s in the various operators involved.
And vacuum subtraction implies regularisation carried out by deducting
the vacuum term that prevails at $t=t_0$ (when $\alpha\,=\,1$ and
$\beta\,=\,0$) from the expectation values.
\smallskip
Though a detailed analysis will be carried out in the next section, to
illustrate the drawbacks of the criterions $\Delta_{KF1}$ and $\Delta_{KF2}$,
we, in this section, evaluate these quantities for a particular case of
the toy model.
When the the quantum system in our toy model is specified to be in a vacuum
state and renormalisation is achieved by normal ordering, the square of the
fluctuations in energy of the quantum oscillator is given by the equation

$$\big\langle\,:{\hat H}^2:\,\big\rangle\;
-\;\big\langle\,:{\hat H}:\,\big\rangle^2\;=\;
{\vert\,\beta\,\vert}^2\;+\;2\,{\vert\,\beta\,\vert}^4.\eqn\qq$$
and the criterions $\Delta_{KF1}$ and $\Delta_{KF2}$ follow the
expressions

$$\Delta_{KF1(NO)}\; =\; \left\lbrace\, {1\; +\; 2\,
{{\vert\, \beta\, \vert}^2} \over {{\vert\, \beta\, \vert}^2}}\,
\right\rbrace\quad;\quad
\Delta_{KF2(NO)}\; =\; \left\lbrace{ 1\; +\; 2\,
{\vert\, \beta\, \vert}^2 \over
{1\;+\; 3\, {\vert\, \beta\, \vert}^2}}\,\right\rbrace.$$
(The subscript $(NO)$ represents normal ordering.)
The above equations clearly show that in the adiabatic limit, {\it i.e}
when $\beta\,\rightarrow\,0$, though the fluctuations in the energy of
the oscillator do vanish completely, the criterions $\Delta_{KF1}$ and
$\Delta_{KF2}$ rather than reducing to zero, as they should, they take on
the values infinity and unity respectively suggesting a breakdown
of the semiclassical theory.
But in the adiabatic limit, the semiclassical theory {\it is} reliable
and the criterions {\it should} vanish.
\smallskip
This `bad' behaviour on the part of these criterions can be corrected
for, if the expectation values of the operators in the quantities
$\Delta_{KF1}$ and $\Delta_{KF2}$ are not renormalised.
Or in other words, if the vacuum terms are re-introduced in the
expectation values that these criterions are made up of.
Performing this, we obtain the criterions necessary for drawing the
limits on the validity of the semiclassical theory for the toy model
to be either the quantity

$$\Delta_{SC1}(t)\; \equiv\; \biggl\vert\,{\big\langle\,{\hat H}^2(t)\,
\big\rangle\; -\; {\big\langle\,{\hat H}(t)\,\big\rangle}^2 \over
{{\big\langle\,{\hat H}(t)\,\big\rangle}^2}}\,\biggl\vert,\eqn\qsconettm$$
(subscript SC stands for semiclassical) or the one

$$\Delta_{SC2}(t)\;\equiv\;\biggl\vert\,{\big\langle\,{\hat H}^2(t)\,\big
\rangle\;-\;{\big\langle\,{\hat H}(t)\,\big\rangle}^2\over
{\big\langle\,{\hat H}^2(t)\,\big\rangle}}\,\biggl\vert.\eqn\qsctwottm$$
The two quantities $\Delta_{SC1}$ and $\Delta_{SC2}$ are related to each
other by the equation

$$\Delta_{SC2}\; =\; \left(\,{\Delta_{SC1} \over {\Delta_{SC1}\; +\; 1}}\,
\right).\eqn\qq$$
When no renormalisation is carried out the square of the fluctuations in the
energy of the oscillator is

$$\big\langle\,{\hat H}^2\,\big\rangle\;
-\;\big\langle\,{\hat H}\,\big\rangle^2\;
=\;2\,{\vert\,\beta\,\vert}^2\;+\;2\,{\vert\,\beta\,\vert}^4$$
so that the quantities $\Delta_{SC1}$ and $\Delta_{SC2}$ are given by the
expressions

$$\Delta_{SC1}\; =\; \left\lbrace\,{2\,{\vert\,\beta\,\vert}^2\; +\; 2\,
{\vert\, \beta\, \vert}^4\over {{\vert\,\beta\,\vert}^2\; +\;
{\vert\, \beta\, \vert}^4\; +{1 \over 4}}}\,\right\rbrace
\quad;\quad\Delta_{SC2}\;  =\;\left\lbrace{2\,{\vert\,\beta\,\vert}^2\;
 +\; 2\,{\vert\, \beta\, \vert}^4\over {3\, {\vert\,\beta\,\vert}^2\;
+\; 3\,{\vert\, \beta\, \vert}^4\; +\; {1 \over 4}}}\,\right\rbrace.$$
The two quantities $\Delta_{SC1}$ and  $\Delta_{SC2}$ do vanish in the
adiabatic limit ({\it i.e} when $\beta\,\rightarrow\,0$), clearly
illustrating that the criterions $\Delta_{SC1}$ and $\Delta_{SC2}$
perform more reliably than either $\Delta_{KF1}$ or $\Delta_{KF2}$.
\bigskip

\beginsection{5. $\Delta_{KF}$ and $\Delta_{SC}$ for different quantum states}

In the following three sub-sections we evaluate the $\Delta$'s for
different states of the quantum oscillator in our toy model.

\medskip
\centerline{\bf (i). In a vacuum state}
\noindent
If the state of the quantum oscillator is defined to be a vacuum state
at the early time $t_0$ (when the other initial conditions have been
specified) then the expectation values of the operators ${\hat H}$ and
${\hat H}^2$ without any renormalisation are

$$\eqalign{E\; =\; \big\langle\, {\hat H}\,\big\rangle
&=\; \big\langle\, 0\,\vert  \left\lbrace\,{\hat a}^{\dagger}
\, {\hat a}\; +\; {1 \over 2}\,\right\rbrace\,\omega
\vert\, 0\,\big\rangle \cr
& =\; \left\lbrace\,{\vert\,\beta\,\vert}^2\; +\; {1 \over 2}
\,\right\rbrace\, \omega \cr}\eqn\qEv$$
and

$$\eqalign{E^2\; =\; \big\langle\,{\hat H}^2\,\big\rangle\;
&=\; \big\langle\, 0\,\vert  \left\lbrace\,
{\hat a}^{\dagger}\, {\hat a}\; +\; {1 \over 2}\,\right\rbrace\,{\omega}\,
\left\lbrace\,{\hat a}^{\dagger}\, {\hat a}\; +\; {1 \over 2}\,\right\rbrace\,
{\omega}\vert\, 0\,\big\rangle\cr
&=\; \left\lbrace\, 3\, {\vert\, \beta\, \vert}^2\; +\; 3\,
{\vert\, \beta\, \vert}^4\;
+\; {1 \over 4}\,\right\rbrace\, \omega^2.\cr}\eqn\qEEv$$
When the operators involved in the above expressions are normal ordered,
the expectation values are

$$\eqalign{E_{(NO)}\; =\; \big\langle\,:{\hat H}:\,\big\rangle
&=\; \big\langle\, 0\,\vert  \left\lbrace\, {\hat a}^{\dagger}
\, {\hat a}\,\right\rbrace\, \omega \vert\, 0\,\big\rangle\cr
& =\; {\vert\,\beta\,\vert}^2\, \omega \cr}\eqn\qq$$
and

$$\eqalign{E_{(NO)}^2\; =\; \big\langle\,:{\hat H}^2:\,\big\rangle\;
&=\; \big\langle\, 0\,\vert
\left\lbrace\,a^{\dagger}\,{\hat a}^{\dagger}\, {\hat a}\,
{\hat a}\,\right\rbrace\,{\omega}^2
\vert\, 0\,\big\rangle\cr
&=\; \left\lbrace\, {\vert\, \beta\, \vert}^2\; +\; 3\,
{\vert\, \beta\, \vert}^4\,\right\rbrace\, \omega^2.\cr}\eqn\qq$$
For the case, when renormalisation is achieved by the vacuum subtraction,
{\it i.e}

$$\eqalign{E_{(VS)}\; &=\; \big\langle\,:{\hat H}:\,\big\rangle\cr
&=\;\big\langle\, 0\,\vert  \left\lbrace\, {\hat a}^{\dagger}
\, {\hat a}\; +\; {1 \over 2}\,\right\rbrace\, \omega \vert\, 0\,
\big\rangle\;-\;
\big\langle\, 0\,\vert  \left\lbrace\, {\hat A}^{\dagger}
\, {\hat A}\; +\; {1 \over 2}\,\right\rbrace\,\omega \vert\, 0\,\big\rangle
\cr}\eqn\qq$$
and

$$\eqalign{E_{(VS)}^2\; &=\; \big\langle\,:{\hat H}^2:\,\big\rangle\cr
&=\;  \big\langle\, 0\,\vert  \left\lbrace\,
{\hat a}^{\dagger}\, {\hat a}\; +\; {1 \over 2}\,\right\rbrace\,\omega\,
\left\lbrace\,{\hat a}^{\dagger}\, {\hat a}\; +\; {1 \over 2}\,
\right\rbrace\,{\omega}
\vert\, 0\,\big\rangle\cr
&-\;\big\langle\, 0\,\vert
\left\lbrace\,
{\hat A}^{\dagger}\, {\hat A}\; +\; {1 \over 2}\,\right\rbrace\,\omega\,
\left\lbrace\,{\hat A}^{\dagger}\, {\hat A}\; +\; {1 \over 2}\,
\right\rbrace\,{\omega}\,
\vert\, 0\,\big\rangle, \cr}\eqn\qq$$
the expressions for $E_{(VS)}$ and $E_{(VS)}$ are the same as the
quantities $E$ and $E^2$ but without the $(\omega/2)$ and the $(\omega^2/4)$
terms respectively.
Substituting the above results in the equations {\qFonetm} and {\qFtwotm},
we obtain that

$$\Delta_{KF1(NO)}\; =\; \left\lbrace\, {1\; +\; 2\,
{{\vert\, \beta\, \vert}^2} \over {{\vert\, \beta\, \vert}^2}}\,
\right\rbrace\quad;\quad
\Delta_{KF2(NO)}\; =\; \left\lbrace{ 1\; +\; 2\,
{\vert\, \beta\, \vert}^2 \over
{1\;+\; 3\, {\vert\, \beta\, \vert}^2}}\,\right\rbrace\eqn\qfonetwono$$
and

$$\Delta_{KF1(VS)}\; =\; \left\lbrace\, {3\; +\; 2\,
{{\vert\, \beta\, \vert}^2} \over {{\vert\, \beta\, \vert}^2}}\,\right
\rbrace\quad;\quad
\Delta_{KF2(VS)}\; =\; \left\lbrace{3\; +\; 2\,
{\vert\, \beta\, \vert}^2 \over
{3\;+\; 3\, {\vert\, \beta\, \vert}^2}}\,\right\rbrace,\eqn\qfonetwovs$$
where the subscripts $(NO)$ and $(VS)$ represent renormalisation
by normal ordering and vacuum subtraction respectively.
\smallskip
{}From the expectation values evaluated above, we find that the criterions
we have suggested for drawing the limits on the semiclassical theory for
the toy model, {\it viz} $\Delta_{SC1\,\&\,2}$ are given by the expressions

$$\Delta_{SC1}\; =\; \left\lbrace\,{2\,{\vert\,\beta\,\vert}^2\; +\; 2\,
{\vert\, \beta\, \vert}^4\over {{\vert\,\beta\,\vert}^2\; +\;
{\vert\, \beta\, \vert}^4\; +{1 \over 4}}}\,\right\rbrace
\quad;\quad\Delta_{SC2}\;  =\;\left\lbrace{2\,{\vert\,\beta\,\vert}^2\;
 +\; 2\,{\vert\, \beta\, \vert}^4\over {3\, {\vert\,\beta\,\vert}^2\;
+\; 3\,{\vert\, \beta\, \vert}^4\; +\; {1 \over 4}}}\,\right\rbrace.
\eqn\qsconetwo$$
\medskip
\centerline{\bf (ii). In a $n$th excited state}
\noindent
For the case when the quantum state of the oscillator is specified to
be a $n$th excited state, the expectation values when no renormalisation
has been carried out are

$$\eqalign{E\; =\; \big\langle\, {\hat H}\,\big\rangle =\;
&=\; \big\langle\, n\,\vert  \left\lbrace\, {\hat a}^{\dagger}
\, {\hat a}\; +\; {1 \over 2}\,\right\rbrace\,\omega \vert\, n\,\big\rangle \cr
& =\; \left\lbrace\,{\vert\,\beta\,\vert}^2\,(2\,n\; +\; 1)\; +\;n\;+\;
{1 \over 2}\,\right\rbrace\, \omega \cr}\eqn\qEn$$
and

$$\eqalign{E^2\; &=\; \big\langle\,:\,{\hat H}^2\,:\,\big\rangle\;
=\; \big\langle\, n\,\vert
\left\lbrace\,{\hat a}^{\dagger}\, {\hat a}\; +\; {1 \over 2}\,\right\rbrace\,
\omega\,
\left\lbrace\,{\hat a}^{\dagger}\, {\hat a}\; +\; {1 \over 2}\,\right\rbrace
\,\omega
\vert\, n\,\big\rangle\cr
&{\eqalign{=\; \biggl\lbrace\, \left(\,n^2\;+ \; n\,\right)\,&\left(\, 1\; +\;
6\, {\vert\, \beta\, \vert}^2\; +\; 6\, {\vert\, \beta\, \vert}^4\,\right)\;\cr
& +\; \left(\,3\, {\vert\, \beta\, \vert}^2
+\; 3\, {\vert\, \beta\, \vert}^4\;+\; {1 \over 4}\,\right)\,\biggl\rbrace\,
\omega^2.}} \cr}\eqn\qEEn$$
When renormalisation is achieved by normal ordering, the expecation values
are given by the expressions

$$\eqalign{E_{(NO)}\; =\; \big\langle\,:{\hat H}:\,\big\rangle
&=\; \big\langle\, n\,\vert  \left\lbrace\, {\hat a}^{\dagger}
\, {\hat a}\,\right\rbrace\,\omega \vert\, n\,\big\rangle \cr
& =\; \left\lbrace\,{\vert\,\beta\,\vert}^2\,\left(2\,n\; +\; 1\,\right)
\; +\;n\,
\right\rbrace\, \omega \cr}\eqn\qq$$
and

$$\eqalign{E_{(NO)}^2\; &=\; \big\langle\,:{\hat H}^2:\,\big\rangle\;
=\; \big\langle\, n\,\vert  \left\lbrace\, {\hat a}^{\dagger}\,
{\hat a}^{\dagger}\, {\hat a}\, {\hat a}\,
\right\rbrace\,{\omega}^2 \vert\, n\,\big\rangle\cr
&{\eqalign{=\; \biggl\lbrace\, n^2\,\left(\, 1\; +\; 6\,
{\vert\, \beta\, \vert}^2
\; +\; 6\, {\vert\, \beta\, \vert}^4\, \right)\; &+\; n\, \left(\,-1\; +\;
2\, {\vert\, \beta\, \vert}^2\; +\; 6\, {\vert\, \beta\, \vert}^4\,\right)\cr
&+\; \left(\, {\vert\, \beta\, \vert}^2\;
+\; 3\; {\vert\, \beta\, \vert}^4\, \right)\, \biggl\rbrace\,
\omega^2.}} \cr}\eqn\qq$$
For regularisation of the expectation values by vacuum subtraction,
{\it i.e}

$$\eqalign{E_{(VS)}\; &=\; \big\langle\,:{\hat H}:\,\big\rangle\cr
&=\; \big\langle\, n\,\vert  \left\lbrace\, {\hat a}^{\dagger}
\, {\hat a}\; +\; {1 \over 2}\,\right\rbrace\,\omega\vert\, n\,\big\rangle
-\;\big\langle\, 0\,\vert  \left\lbrace\, {\hat A}^{\dagger}
\, {\hat A}\; +\; {1 \over 2}\,\right\rbrace\,\omega\vert\, 0\,\big\rangle
\cr}\eqn\qq$$
and

$$\eqalign{E_{(VS)}^2\; &=\; \big\langle\,:{\hat H}^2:\,\big\rangle\cr
&=\; \big\langle\, n\,\vert  \left\lbrace\,
{\hat a}^{\dagger}\, {\hat a}\; +\; {1 \over 2}\right\rbrace\,\omega\,
\left\lbrace\,{\hat a}^{\dagger}\, {\hat a}\; +\; {1 \over 2}\,\right\rbrace
\,\omega
\vert\, n\,\big\rangle\cr
&-\;\big\langle\, 0 \,\vert\left\lbrace\,{\hat A}^{\dagger}\,
{\hat A}\; +\; {1 \over 2}\,\right\rbrace\,\omega\,
\left\lbrace\,{\hat A}^{\dagger}\, {\hat A}\; +\; {1 \over 2}\,\right
\rbrace\,\omega \vert\, 0\,\big\rangle, \cr}\eqn\qq$$
the expressions for $E_{(VS)}$ and $E_{(VS)}^2$ are the same as
the quantities $E$ and $E^2$ but without the $\,({\omega/2})\,$ and
$\,({\omega^2/4})\,$ terms respectively.
Substituting the quantities evaluated above in the equations {\qFonetm}
and {\qFtwotm}, we find that the criterions suggested by Kuo and Ford
are given by the following expressions:

$$\Delta_{KF1(NO)}\;=\; \left\lbrace\, \biggl\vert\, {{\vert\,\beta\,\vert}^4\,
\left(\,2\, n^2\; +\; 2\,n \;+\; 2\,\right)\; +\; {\vert\, \beta\, \vert}^2\,
\left(\,2\,n\; +\; 1\,\right)\; -\; n \over {{\vert\, \beta\, \vert}^4\,
\left(\,4\, n^2\; +\; 4\,n \;+\; 1\,\right)\; +\; {\vert\, \beta\, \vert}^2\,
\left(\,4\,n^2\; +\; 2\,n\,\right)\; +\; n^2}}\, \biggl\vert\,
\right\rbrace\eqn\qq$$

$$\Delta_{KF2(NO)}\;=\;\left\lbrace\, \biggl\vert\, {{\vert\,\beta\,\vert}^4\,
\left(\,2\, n^2\; +\; 2\,n \;+\; 2\,\right)\; +\; {\vert\,\beta\, \vert}^2\,
\left(\,2\,n\; +\; 1\,\right)\; -\; n \over {{\vert\,\beta\, \vert}^4\,
\left(\,6\, n^2\; +\; 6\,n \;+\; 3\,\right)\; +\; {\vert\, \beta\, \vert}^2\,
\left(\,6\,n^2\; +\; 2\,n\; +\; 1\,\right)\; +\; (n^2\; -\; n)}}\,
\biggl\vert\,\right\rbrace\eqn\qq$$

$$\Delta_{KF1(VS)}\;=\; \left\lbrace\, {\vert\, \beta\, \vert}^4\,
\left(\,2\,n^2\; +\; 2\,n \;+\; 2\,\right)\; +\; {\vert\, \beta\, \vert}^2\,
\left(\,2\,n^2\; +\; 4\,n\; +\; 3\,\right)\;+\; n \over
{{\vert\,\beta\,\vert}^4\,
\left(\,4\, n^2\; +\; 4\,n \;+\; 1\,\right)\;+\; {\vert\, \beta\,\vert}^2\,
\left(\,4\, n^2\;+\;2\,n\,\right)\;+\;n^2 }\,\right\rbrace\eqn\qq$$
and

$$\Delta_{KF2(VS)}\;=\; \left\lbrace\, {\vert\, \beta\, \vert}^4\,
\left(\,2\, n^2\; +\; 2\,n \;+\; 2\,\right)\; +\; {\vert\,\beta\,\vert}^2\,
\left(\,2\,n^2\; +\; 4\,n\; +\; 3\,\right)\; +\; n \over {\left(\,{\vert\,
 \beta \,\vert}^4\;
+\;{\vert\,\beta\,\vert}^2\,\right)\,\left(\,6\, n^2\; +\; 6
\,n \;+\; 3\,\right)\;
+\;\left(\,n^2\;+\;n\,\right)}\, \right\rbrace.\eqn\qq$$
Whereas the criterions $\Delta_{SC1\,\&\,2}$, when the expectation values
$E$ and $E^2$ are substituted in the equations {\qsconettm} and
{\qsctwottm} are given by the expressions

$$\Delta_{SC1}\; =
\; \left\lbrace\, \left(\,{2\, {\vert\, \beta \, \vert}^2\;
+\; 2\,{\vert\, \beta \, \vert}^4 \over {\left(\,1\; +\; 2\,
{\vert\, \beta \, \vert}^2\,\right)^2}}\,\right)\, \left(\,{n^2\; +n\;
+ 1 \over {n^2\; +n\; +\; {1 \over 4}}}\,\right)\,\right\rbrace\eqn\qq$$
and

$$\Delta_{SC2}\; = \;
\left\lbrace\, \left(\,{2\, {\vert\, \beta\, \vert}^2\; +\; 2
\,{\vert\, \beta \, \vert}^4 \over
{1\; +\; \; 6\,{\vert\, \beta\, \vert}^2\,
+\; 6\, {\vert\, \beta\, \vert}^4}}\right)\,
\left(\,{n^2\; +n\; + 1 \over {n^2\; +n\;
+\; {1 \over 2}}}\,\right)\,\right\rbrace.\eqn\qq$$
\medskip
\centerline{\bf (iii). In a  coherent state}
\noindent
When the quantum oscillator is specified to be in a coherent state
the expectation values when no renormalisation is carried out are
given by the equations

$$\eqalign{E\; &=\; \big\langle\,{\hat H}\,\big\rangle
=\; \big\langle\, \lambda\, \vert  \left\lbrace\, {\hat a}^{\dagger}
\, {\hat a}\; +\; {1 \over 2}\,\right\rbrace\,\omega
\vert\, \lambda\, \big\rangle \cr
& =\; \left\lbrace\, {\vert\, \lambda\, \vert}^2\,
\left(\, 1\; +\; 2\, {\vert\,\beta\,\vert}^2\right)\; +\; {\lambda}^2\,
\alpha\, \beta\; +\; {{\lambda}^*}^2\, {\alpha}^*\, {\beta}^*\; +\;
{\vert\, \beta\, \vert}^2\; +\; {1 \over 2}\,\right\rbrace\,
\omega \cr}\eqn\qq$$
and

$$\eqalign{E^2\; &=\; \big\langle\,{\hat H}^2\,\big\rangle\;
=\; \big\langle\, \lambda \,\vert  \left\lbrace\,{\hat a}^{\dagger}\,
{\hat a}\; +\; {1 \over 2}\,\right\rbrace\,\omega
\left\lbrace\,{\hat a}^{\dagger}\, {\hat a}\; +\; {1 \over 2}\,
\right\rbrace\,\omega\vert \lambda \big\rangle\cr
&=\; \biggl\lbrace\, \left(\,{\vert\, \lambda\, \vert}^4\; +\;
2\, {\vert\, \lambda\, \vert}^2\,\right)\, \left(\, 1\; +\;
6\, {\vert\, \beta \, \vert}^2\; +\; 6\, {\vert\,\beta\,\vert}^4\,
\right)\cr
&+\; \left(\, 2\, {\vert\, \lambda\, \vert}^2\; +\; 3\,\right)\,
\left({\lambda}^2\,\alpha\,\beta\; +\; {\lambda^*}^2\,
\alpha^*\,\beta^*\,\right)\, \left(\,1\; +\; 2\, {\vert\,\beta\,\vert}^2\,
\right)\cr
&+\; \left(\,{\lambda}^4\,{\alpha}^2\,{\beta}^2\;
+\; {\lambda^*}^4\,{\alpha^*}^2\,{\beta^*}^2\, \right)\cr
&+\; \left(\, 3{\vert\,\beta \,\vert}^2\; +\; 3\,{\vert\, \beta\, \vert}^4\,
\right)\; +\; {1 \over 4}\, \biggl\rbrace\,
\omega^2. \cr}\eqn\qq$$
When the operators are normal ordered the expectation values are

$$\eqalign{E_{NO}\; &=\; \big\langle\,:{\hat H}:\,\big\rangle
=\; \big\langle\, \lambda\, \vert  \left\lbrace\, {\hat a}^{\dagger}
\, {\hat a}(t)\,\right\rbrace\,\omega
\vert\, \lambda\, \big\rangle \cr
& =\; \left\lbrace\, {\vert\, \lambda\, \vert}^2\,
\left(\, 1\; +\; 2\,{\vert\,\beta\,\vert}^2\right)\; +\; {\lambda}^2\,
\alpha\, \beta\; +\; {{\lambda}^*}^2\, {\alpha}^*\, {\beta}^*\; +\;
{\vert\, \beta\, \vert}^2\,\right\rbrace\,
\omega \cr}\eqn\qq$$
and

$$\eqalign{E^2_{NO}\; &=\; \big\langle\,:{\hat H}^2:\,\big\rangle\;
=\; \big\langle\, \lambda \,\vert  \left\lbrace\,{\hat a}^{\dagger}\,
{\hat a}^{\dagger}\,{\hat a}\, {\hat a}\,
\right\rbrace\,{\omega}^2\vert \lambda \big\rangle\cr
&=\; \biggl\lbrace\, {\vert\, \lambda\,\vert}^4\,
\left(\, 1\; +\; 6\, {\vert\, \beta \,\vert}^2\;+\;
6\,{\vert\, \beta\, \vert}^4\, \right)\cr
&+\;{\vert\,\lambda \,\vert}^2\,\left(\,
8\, {\vert\, \beta\,\vert}^2\;+\; 12\,{\vert\, \beta\,\vert}^4\,\right)\cr
&+\; \left(\,{\lambda}^2\, \alpha\,\beta\;
+\; {\lambda^*}^2\, {\alpha}^*\, {\beta}^*\,\right)\,
\left\lbrace\,1\; +\; 6\, {\vert\, \beta\, \vert}^2\; +\;
{\vert \, \lambda \, \vert}^2\, \left(\, 2\;
+\; 4\, {\vert\, \beta\, \vert}^2\,\right)\,\right\rbrace\cr
&+\; \left({\lambda}^4\, {\alpha}^2\, {\beta}^2\;
+\; {\lambda^*}^4\, {\alpha^*}^2\, {\beta^*}^2\,\right)
+\; 3\, {\vert\, \beta\, \vert}^4\; +\; {\vert\, \beta\,\vert}^2\,
\biggl\rbrace\,
{\omega}^2. \cr}\eqn\qq$$
For the case, when renormalisation is achieved by vacuum subtraction,
{\it i.e}

$$\eqalign{E_{(VS)}\; &=\; \big\langle\,:{\hat H}:\,\big\rangle\cr
&=\; \big\langle\, \lambda \,\vert  \left\lbrace\, {\hat a}^{\dagger}
\, {\hat a}\; +\; {1 \over 2}\,\right\rbrace\,\omega\vert\,
\lambda \,\big\rangle
-\;\big\langle\, 0\,\vert  \left\lbrace\, {\hat A}^{\dagger}
\, {\hat A}\; +\; {1 \over 2}\,\right\rbrace\,\omega\vert\, 0\,\big\rangle
\cr}\eqn\qq$$
and

$$\eqalign{E_{(VS)}^2\; &=\; \big\langle\,:{\hat H}^2:\,\big\rangle\cr
&=\; \big\langle\, \lambda\,\vert  \left\lbrace\,
{\hat a}^{\dagger}\, {\hat a}\; +\; {1 \over 2}\right\rbrace\,\omega\,
\left\lbrace\,{\hat a}^{\dagger}\, {\hat a}\; +\; {1 \over 2}\,\right\rbrace
\,\omega
\vert\, \lambda\,\big\rangle\cr
&-\;\big\langle\, 0 \,\vert\left\lbrace\,{\hat A}^{\dagger}\,
{\hat A}\; +\; {1 \over 2}\,\right\rbrace\,\omega\,
\left\lbrace\,{\hat A}^{\dagger}\, {\hat A}\; +\; {1 \over 2}\,\right
\rbrace\,\omega \vert\, 0\,\big\rangle, \cr}\eqn\qq$$
the expectation values, $E_{(VS)}$ and $E_{(VS)}^2$ are given by the same
expressions as the quantities $E$ and $E^2$ but without the $(\omega/2)$
and $(\omega^2/4)$ terms respectively.
For the coherent state being considered, the expressions for the
$\Delta$'s prove to be rather lenghthy.
Due to this reason, we do not write them down here explicitly but
just quote their values in the different limits of interest in the
tables below.
\smallskip
The numerators of the various $\Delta$'s evaluated earlier in this section
contain either the quantity
$\left(\,\big\langle\,:{\hat H}^2:\,\big\rangle\;
-\;\big\langle\,:{\hat H}:\,\big\rangle^2\,\right)$
or the one
$\left(\,\big\langle\,{\hat H}^2\,\big\rangle\;
-\;\big\langle\,{\hat H}\,\big\rangle^2\,\right)$,
both of them being the square of the fluctuations in the energy of the
quantum oscillator in the toy model.
These quantities, as can be seen from the expressions for the $\Delta$'s,
are proportional to at least the second power of $\vert\,\beta\,\vert$
and hence a large value for $\beta$ would imply large fluctuations and
hence a breakdown of the semiclassical theory.
\smallskip
The expressions for the different $\Delta$'s in the two limits of interest,
{\it viz} $\beta\,\rightarrow\,0$ and $\beta\,\rightarrow\,\infty$ are
summarised in the tables I and II respectively.
\bigskip
\bigskip
\centerline{\bf Table I ($\beta\,\rightarrow\,0$)}
\medskip
\settabs 4\columns
\hrule
\hrule
\medskip
\+{}&{\hfill Vacuum \hfill}&{\hfill $n$th excited \hfill}
&{\hfill \hfill Coherent \hfill}&{}\cr
\medskip
\hrule
\hrule
\bigskip
\+{\hfill $\Delta_{SC1}$ \hfill}&{\hfill $0$ \hfill}
&{\hfill $0$ \hfill}
&{\hfill $\left(\,{{\vert\, \lambda\, \vert}^2 \over
{{\vert\,\lambda\,\vert}^4\;+\;{\vert\,\lambda\,\vert}^2\;
+\;{1\over 4}}}\,\right)$ \hfill}&{}\cr
\bigskip
\+{\hfill $\Delta_{SC2}$ \hfill}&{\hfill $0$ \hfill}&{\hfill $0$ \hfill}
&{\hfill $\left(\,{\vert\, \lambda\, \vert^2 \over
{{\vert\,\lambda\,\vert}^4\;+\;2\,{\vert\,\lambda\,\vert}^2\;
+\;{1\over 4}}}\,\right)$ \hfill}&{}\cr
\bigskip
\+{\hfill $\Delta_{KF1(NO)}$ \hfill}&{\hfill $\infty$ \hfill}
&{\hfill $\left(\,{1 \over n}\,\right)$ \hfill}
&{\hfill $0$ \hfill}&{}\cr
\bigskip
\+{\hfill $\Delta_{KF2(NO)}$ \hfill}&{\hfill $1$ \hfill}
&{\hfill $\left(\,{1 \over n\;-\;1}\,\right)$ \hfill}
&{\hfill $0$ \hfill}&{}\cr
\bigskip
\+{\hfill $\Delta_{KF1(VS)}$ \hfill}&{\hfill $\infty$ \hfill}
&{\hfill $\left(\,{1 \over n}\,\right)$ \hfill}
&{\hfill $\left(\,{2 \over {\vert\,\lambda\,\vert}^2}\,\right)$ \hfill}&{}\cr
\bigskip
\+{\hfill $\Delta_{KF2(VS)}$ \hfill}&{\hfill $1$ \hfill}
&{\hfill $\left(\,{1 \over n\;-\;1}\,\right)$ \hfill}
&{\hfill $\left(\,{2 \over {2\;+\;{\vert\,\lambda\,\vert}^2}}\,\right)$
\hfill}&{}\cr
\bigskip
\hrule
\hrule
\vfill\eject
\centerline{\bf Table II ($\beta\,\rightarrow\,\infty$)}
\medskip
\settabs 6\columns
\hrule
\hrule
\medskip
\+{}&{\hfill Vacuum \hfill}&{\hfill \quad $n$th excited \hfill}
&{}&{\hfill Coherent  \quad \hfill}&{}\cr
\medskip
\hrule
\hrule
\bigskip
\+{$\Delta_{SC1}$}&{\hfill $2$ \hfill}
&{\hfill $\left(\,{n^2\;+\;n\;+\;1 \over {2\,n^2\;+\;2\,n\;
+\;{1\over 2}}}\,\right)$ \hfill}
&{\qquad \qquad $\left(\,{{\vert\,\lambda\,\vert}^2\,\left(\,8\;
+\;4\,c_1\,\right)\;+\;2 \over {\left(\,{\vert\,\lambda\,\vert}^2\,
\left(\,2\;+\;c_1\,\right)\;+\;1\,\right)^2}}\,\right)$}
&{}\cr
\bigskip
\+{$\Delta_{SC2}$}&{\hfill $\left(\,2 \over 3\,\right)$ \hfill}
&{\hfill $\left(\,{n^2\;+\;n\;+\;1 \over {3\,n^2\;+\;3\,n\;
+\;{3\over 2}}}\,\right)$ \hfill}
&{\quad $\left(\,{{\vert\,\lambda\,\vert}^2\,\left(\,8\;
+\;4\,c_1\,\right)\;+\;2 \over {{\vert\,\lambda\,\vert}^4\,
\left(\,6\;+\;4\,c_1\;+\;c_2\,\right)\;+\; {\vert\,\lambda\,\vert}^2\,
\left(\,12\;+\;6\,c_1\,\right)\;+\;3}}\,\right)$ \hfill}&{}&{}\cr
\bigskip
\+{$\Delta_{KF1(NO)}$}&{\hfill $2$ \hfill}
&{\hfill $\left(\,{n^2\;+\;n\;+\;1 \over {2\,n^2\;+\;2\,n\;
+\;{1\over 2}}}\,\right)$ \hfill}
&{\qquad \qquad $\left(\,{{\vert\,\lambda\,\vert}^2\,\left(\,8\;
+\;4\,c_1\,\right)\;+\;2 \over {\left(\,{\vert\,\lambda\,\vert}^2\,
\left(\,2\;+\;c_1\,\right)\;+\;1\,\right)^2}}\,\right)$}
&{}\cr
\bigskip
\+{$\Delta_{KF2(NO)}$}&{\hfill $\left(\,2 \over 3\,\right)$ \hfill}
&{\hfill $\left(\,{n^2\;+\;n\;+\;1 \over {3\,n^2\;+\;3\,n\;
+\;{3\over 2}}}\,\right)$ \hfill}
&{\quad $\left(\,{{\vert\,\lambda\,\vert}^2\,\left(\,8\;
+\;4\,c_1\,\right)\;+\;2 \over {{\vert\,\lambda\,\vert}^4\,
\left(\,6\;+\;4\,c_1\;+\;c_2\,\right)\;+\; {\vert\,\lambda\,\vert}^2\,
\left(\,12\;+\;6\,c_1\,\right)\;+\;3}}\,\right)$ \hfill}&{}&{}\cr
\bigskip
\+{$\Delta_{KF1(VS)}$}&{\hfill $2$ \hfill}
&{\hfill $\left(\,{n^2\;+\;n\;+\;1 \over {2\,n^2\;+\;2\,n\;
+\;{1\over 2}}}\,\right)$ \hfill}
&{\qquad \qquad $\left(\,{{\vert\,\lambda\,\vert}^2\,\left(\,8\;
+\;4\,c_1\,\right)\;+\;2 \over {\left(\,{\vert\,\lambda\,\vert}^2\,
\left(\,2\;+\;c_1\,\right)\;+\;1\,\right)^2}}\,\right)$}
&{}\cr
\bigskip
\+{$\Delta_{KF2(VS)}$}&{\hfill $\left(\,2 \over 3\,\right)$ \hfill}
&{\hfill $\left(\,{n^2\;+\;n\;+\;1 \over {3\,n^2\;+\;3\,n\;
+\;{3\over 2}}}\,\right)$ \hfill}
&{\quad $\left(\,{{\vert\,\lambda\,\vert}^2\,\left(\,8\;
+\;4\,c_1\,\right)\;+\;2 \over {{\vert\,\lambda\,\vert}^4\,
\left(\,6\;+\;4\,c_1\;+\;c_2\,\right)\;+\; {\vert\,\lambda\,\vert}^2\,
\left(\,12\;+\;6\,c_1\,\right)\;+\;3}}\,\right)$ \hfill}&{}&{}\cr
\bigskip
\hrule
\hrule
\bigskip
\bigskip
\noindent
The quantities $c_1$ and $c_2$ in the table II are given to be

$$c_1\;=\;2\, \cos\,(\,a\;+\;b\;+\;2\,l\,)\quad;\quad
c_1\;=\;2\, \cos\,(\,2\,a\;+\;2\,b\;+\;4\,l\,)\eqn\qq$$
where $a$, $b$ and $l$ are the arguments of the complex quantities
$\alpha$, $\beta$ and $\lambda$ respectively.
\smallskip
As listed in Table I, in the adiabatic limit, the quantities $\Delta_{SC1}$
or $\Delta_{SC2}$ do not vanish for the coherent state because these states
are not energy eigen states and hence they do possess fluctuations in the
energy.
Whereas, when the operators are normal ordered these fluctuations in the
energy for the coherent state do vanish in the adiabatic limit, as is
reflected by the values of $\Delta_{KF1}$ and $\Delta_{KF2}$ in Table I.
For the other vacuum and the $n$th excited states the quantities
$\Delta_{KF1}$ and $\Delta_{KF2}$ do not vanish in the adiabtic limit
whereas the criterions $\Delta_{SC1}$ or $\Delta_{SC2}$ do and hence,
as mentioned in the previous section, perform more reliably.
So, though all the criterions listed in tables above yield equivalent
results  in the limit $\beta\,\rightarrow\,\infty$, to draw the limits
on the validity of the semiclassical theory for the toy model we have to
concentrate on the quantities $\Delta_{SC1}$ or $\Delta_{SC2}$.
\smallskip
In the limit when $\beta$ is large, it can be seen from Table II, that
the criterions  $\Delta_{SC1}$ and $\Delta_{SC2}$ are of order unity
for the vacuum and the $n$th excited state.
Hence the semiclassical theory will not prove to be reliable if the
quantum oscillator is specified to be in one of these two states.
Excited states with a large value for $n$ are generally assumed to be
reliable states to study the semiclassical theory.
In the limit $\beta\,\rightarrow\,\infty$ the criterions $\Delta_{SC1}$
and $\Delta_{SC2}$ are of order unity, in $n$th excited states for
large $n$, suggesting the breakdown of the theory.
\smallskip
Whereas, if the quantum oscillator is specified to be in a coherent state,
the results tabulated for the criterions $\Delta_{SC1}$ and $\Delta_{SC2}$,
clearly show that, for a large value for the parameter $\lambda$ of the
coherent state, thses dimensionless criterions die down as
$\left(\,1\,/\,{\vert\,\lambda\,\vert}^2\,\right)$ irrespective of the
value of $\beta$.
So the semiclassical theory being considered for the toy model will
prove to be absolutely reliable if such states are specified for the
quantum oscillator.
\bigskip
\beginsection{6. Conclusions}

The results of the above section quite clearly prove that the semiclassical
theory being considered for the toy model, can be relied upon, during
all stages of the evolution, only if the quantum system is specified to be
in coherent like states.
It is quite possible, due to the nature of the coupling between the
degrees of feedom chosen for the toy model, that this conclusion might
prove to be valid even for quantised scalar fields in time dependant metrics.
If the backreaction problem {\it has} to be in studied in states,
for the quantum system, which do not possess a coherent nature, the
semiclassical theory being considered in this paper is bound to prove
rather inadequate and the fluctuations will have to be accounted for
in the backreaction term.
When done so, the semiclassical theory can be expected to be described by
an equation similar in form to the Langevin equation.
\medskip
It has been claimed, in section {\bf 4} and the discussions following
the tables in section {\bf 5}, that the criterions $\Delta_{SC1}$ and
$\Delta_{SC2}$ perform more reliably than either $\Delta_{KF1}$ or
$\Delta_{KF2}$.
This was achieved by introducing the vacuum terms in the criterions
$\Delta_{KF1}$
and $\Delta_{KF2}$ to yield the quantities $\Delta_{SC1}$
and $\Delta_{SC2}$ so that they provide reliable results in the adiabatic
limit.
For the case of quantum field theory, the quantities $\Delta_{SC1}$ and
$\Delta_{SC2}$ which involve non-renormalised expectation values, will
prove to be a ratio of divergences and hence will make no sense.
So, only quantities involving renormalised expectation values can be
evaluated.
Since, it has been illustrated, for the case of the toy model, that the
quantities $\Delta_{KF1}$ and $\Delta_{KF2}$ prove to be unreliable in
the adiabatic limit, for the field theoretic case, it would be advisable
to concentrate on just the fluctuations in the energy-momentum density of
the quantum field.
Negligible fluctuations can then be considered to be a positive result for
the validity of the semiclassical theory.
And, when there is a prolific production of particles taking place, either
$\Delta_{KF1}$ or $\Delta_{KF2}$ can be relied upon to reflect the validity
of the semiclassical theory.
\smallskip
Though, in this paper, the unreliability of the criterions $\Delta_{KF1}$
and $\Delta_{KF2}$ in the adiabatic limit and means of improving upon
these quantities was pointed out for the case of the toy model, its main
objective was to illustrate the limited validity of the semiclassical theory.
\bigskip
\beginsection
\centerline{\bf Acknowledgements}

\noindent
The author is being supported by the Senior Research Fellowship of the
Council of Scientific and Industrial Research, India. This work was
done under the guidance of T.~Padmanabhan.
\bigskip
\beginsection
\centerline{\bf References}

\medskip
\noindent
[1] B. S. DeWitt, Phys. Rep. C {\bf 19}, 295 (1975)
\medskip
\noindent
[2] V. G. Lapchinsky and V. A. Rubakov, Acta Phys. Polon. B {\bf 10},
1041 (1975)
\medskip
\noindent
[3] J. B. Hartle, in {\sl Gravitation in Astrophysics} (J. B. Hartle
and B. Carter, Eds.), Plenum, New York (1986).

\medskip
\noindent
[4] T. Padmanabhan and T. P. Singh, Ann. Phys. (N.Y.) {\bf 221}, 217 (1993).
\medskip
\noindent
[5] J. J. Sakurai, {\sl Modern Quantum Mechanics}, Addison Wesley (1985).
\medskip
\noindent
[6] Ya. B. Zeldovich and A. A. Starobinskii, ZhETF {\bf 61}, 2161 (1971).
\medskip
\noindent
[7] Chung-I Kuo and L. H. Ford, Phys. Rev. D {\bf 47}, 4510 (1993).

\end